\documentclass[12pt]{article}
\usepackage{a4,amsmath,amssymb}

\newcommand{\ben}{\begin{eqnarray}}
\newcommand{\een}{\end{eqnarray}}

\newcommand{\half}{\frac{1}{2}}

\begin{document}
\title{Point Electric Charge in General Relativity}
\author{P.~P.~Fiziev\footnote{ E-mail:\,\, fiziev@phys.uni-sofia.bg}, ~S.~V.~Dimitrov
\footnote{E-mail:\,\, stanimir@phys.uni-sofia.bg}} \maketitle
\begin{center}
{\it Department of Theoretical Physics, Faculty of Physics, Sofia University}\\
{\it 5 James Bourchier Boulevard, 1164 Sofia, Bulgaria.}
\end{center}
\vskip 1.0cm \hrule
\begin{abstract}
\begin{small}
\noindent Using a proper gauge condition the static spherically symmetric solutions of Einstein-Maxwell equations 
with charged point source at the center are derived. It is shown that the solutions of the field equations are a 
three-parameter family depending on the Keplerian mass $M$, the charge $Q$ and the bare mass $M_0$. The result can 
be interpreted as a correction to Newton's gravitational potential and Coulomb's electric potential which are both 
regular at the centre where the massive point is placed. A correction to Gauss theorem is derived based on the nontrivial 
topology of the corresponding spacetime.
\end{small}
\end{abstract}
\hrule\vskip 1.0cm 
%
\setcounter{footnote}{0}
\section{Introduction}
\maketitle
\setcounter{equation}{0}
Since the beginning of the 20th century just after the proposal of 
General Relativity (GR) there were several attempts to describe the gravitational 
field of a physical point mass. It was a problem with interesting 
consequences in Maxwell electrodynamics and Newtonian gravity. The problem 
was that Maxwell electrodynamics implies relativity principles while 
in case of Newton gravity they did not hold. After GR was invented the simplest 
solutions considering spherical symmetry and vacuum conditions were 
discovered. These are directly connected to the physical point source outside the 
source due to symmetry reasons but right after their discovery a new problem 
appeared. The vacuum solution of GR did not fit into the new for that time 
mathematical description of a physical point source. The latter was worked out 
by Paul Dirac who proposed a new class of unusual ``functions'' describing the point 
source and obtained as a limit of some real density distributions. They were called 
distributions (generalized functions) and have quite nontrivial mathematical properties 
\cite{Schwartz}, \cite{Gelfand}, \cite{Bremer}. These functions are strongly connected to the physical definition of 
a point source which is quite different from the mathematical one. Thus in order to 
understand the point source problem its the very definition that has to be considered.

Coming from the time of Newton there is a significant difference between the meaning of 
geometrical point in mathematics and point mass or point charge in physics. First of all the abstract 
mathematical definition of a geometrical point is related to some dimensional properties, measure, etc. 
of this object when considered in the presence of other points, i.e. it is a zero dimensional 
object satisfying some extra assumptions when put into a set of other geometrical points. The corresponding 
physical definition is rather different instead. It rises when considering a finite object with 
well defined dimensions, for example it is a sphere of radius $a$, a cube with side length 
$a$ or whatever else. This object is considered as a physical point only if an observer is placed 
in the picture at a distance $L$ and the condition $L\gg a$ holds. This definition strongly depends 
on whether such condition holds and leads to some interesting effects when point source solutions are 
considered. For example, in classical electrodynamics when static solutions are taken into account, the 
electric field $\varphi$ of such a physical point charge is given by the simple relation $\varphi\sim1/L$. 
If one tries to use this solution when the point source condition is not valid it is obvious that there 
is a significant divergence of the field. The question rising at this point is whether such solutions can 
be used when $L\sim a$ or even when $L\ll a$. The answer is well known since the invention of quantum 
electrodynamics (QED) -- NO. It is because in such cases the physical structure of the "point" starts 
playing a significant role and one has to impose some quantum corrections to the electric potential.

In case of GR however there were many attempts to derive a description of such a point source but none of 
them was satisfactory. In a recent paper \cite{fiziev} a way of finding such solution was described. It differs 
from previous attempts in the assumption that the well known gauge freedom of GR was used to regularize Einstein 
field equations. In many of the previous solutions this gauge freedom was neglected due to coordinate independence 
of all physical theories. However it is a well known mathematical fact that the use of {\em singular} transformations 
may lead to contradiction in the theory and thus has to be avoided. This is the reason for choosing only {\em regular} 
coordinate transformations in order to arrive at a correct theory. These regularity conditions in case of point mass 
lead to very interesting new effects. The main result is that the vacuum solution was cut to some value outside the 
coordinate dependant event horizon thus leading to a finite luminosity of the point mass in a very natural way. 
In other words this solution provided a regularization of the gravitational field of the physical point source 
and a corresponding correction to Newton's law typically acting only when approaching the gravitational (Schwarzschild) radius. 

Since the case of a point mass was regular it was interesting to try to solve the corresponding problem when 
the point is charged. Thus besides its gravitational field it produces also electric field in the static case. 
This article is devoted to derivation of such solution and the results are even more astonishing than the previous. 
First of all this method for the first time allows the description of real elementarr particles as well as astrophysical 
objects. Also besides the regular behaviour of the gravitational field it turns out that gravity regularizes in a natural 
way the electric field as well. There is also corrections to both Newton's and Coulomb's laws where the second is regularized 
at distances near the classical radius of the point source in case of elementary particle like solutions. Thus the interplay 
between gravity and electrodynamics opens new unexpected ways for regularization of classical divergencies and 
description of finite classical (not quantum) models of elementary particles. If this concept is right these classical 
corrections together with the proved to be finite quantum ones may lead to a new unified description of elementary 
particles.

Having in mind the fundamental character of these considerations their detailed derivation and discussion will be 
delivered in the present article.
\section{Action Principle and Field Equations}
In order to obtain solutions of Einstein-Maxwell equations in case of a point source 
it is the spherical symmetry and the static character of spacetime that will be the 
starting point. It is well known that each symmetry of the action integral for a 
particular system reduces the degrees of freedom thus leading to a specific kind of 
dimensional reduction.

In case of spherical symmetry in GR it is the group $SO(3)$ that acts invariantly as an 
isometry on the Riemann spacetime manifold $\mathbb{M}^{(1,3)}\{g_{ij}(x)\}$. In 
\cite{moderngeometry} this invariance of the action integral was used to demonstrate 
that the latter can be reduced to two dimensional action represented as an integral with 
respect to the cosmic time $t$ and radial coordinate $r$ on the orbit space 
$\mathbb{M}^{(1,1)}=\mathbb{M}^{(1,3)}\{g_{ij}(x)\}/SO(3)$. This reduction is a consequence 
only of the spherical symmetry of the problem.

Considering static solutions of Einstein-Maxwell equations in the rest frame of the 
source there is another symmetry of spacetime -- it is invariant under one dimensional translations 
$T(1)$ with respect to the cosmic time $t$. This additional symmetry reduces the problem to one 
dimensional so that it is described in terms of a single essential space-time coordinate -- the 
radial variable $r\in [0,\infty)$ as a proper coordinate on the one--dimensional manifold 
$\mathbb{M}^{(1)}=\mathbb{M}^{(1,1)}/T(1)$.

The reduction of the classical action due to spherical symmetry is not the usual way 
to solve Einstein--Maxwell equations for a point particle but it will be the one used here. 
Such procedure was first introduced by Weyl and discussed by Levi--Civita \cite{Weyl} 
later. It yields the same solutions as the direct use of Einstein equations and is 
based on restriction of the classical action on the orbits of the $SO(3)$ group. The 
reason for choosing this type of description is that the action principle gives the 
simplest opportunity for solving the field equations correctly with their invariant 
distributions on the right hand side. These distributions arise from considering the 
particle's mass and charge density as proportional to the three dimensional invariant Dirac 
delta functions.
\subsection{Reduction of the Field's Action Integrals}
The most general form of the spherically symmetric line element is described by three 
independent functions in the metric tensor:
\ben\label{metric} ds^2=g_{\text{tt}}(r)dt^2+g_{\text{rr}}(r)dr^2-\rho(r)^2d\theta^2-\rho(r)^2\sin^2{\theta}d\phi^2. 
\een 
According to Birgkhoff's theorem \cite{moderngeometry}, \cite{gravity} the spacetime is static so 
that the metric components $g_{\text{tt}}>0$, $g_{\text{rr}}<0$ and $\rho>0$ are unknown 
functions of the radial coordinate $r$ only.

Because of the gauge freedom in GR the gravitational field depends on only two of these 
components so that one of them can be fixed as a function of $r$. From geometrical point 
of view one has to introduce an additional mathematical structure describing some specific 
embedding of the manifold $\mathbb{M}^{(1)}$ into the spacetime $\mathbb{M}^{(1,3)}$. This 
can be done in different ways. If {\em regular} gauge transformations are used, the results 
do not depend on the choice of this embedding, i.e. the regular gauge freedom appears. However the 
use of {\em singular} gauge transformations can change the physical properties of the 
solutions, see \cite{fiziev} and the references therein. The most common gauge condition in 
solving Einstein equations is $\rho(r)=r$ which was first used by Hilbert and provides the 
well known Reissner-Nordstr$\ddot{\text{o}}$m form of the vacuum solution of Einstein-Maxwell 
equations. In the present article this gauge freedom will be kept until a proper condition 
for gauge fixing is derived. The proper condition comes from regularity of the field 
equations, i.e. finite multiplier of the delta functions.
\subsubsection{Gravitational Field Lagrangian}
In order to reduce the action for the gravitational field without fixing the gauge proper 
second derivatives have to be extracted from the Ricci scalar. Then the Einstein-Hilbert part 
of the action, according to \cite{fiziev}, \cite{fock}, gives: 
\ben 
\nonumber &&\mathcal{A}_{GR}=\frac{1}{16\pi}\int_{\mathbb{M}^{(1,3)}}d^4x\sqrt{|g^{(4)}|}R= 
\int_{\mathbb{M}^{(1,1)}_{t,r}}dt\,dr\mathcal{L}_{GR}, 
\een 
where $\mathcal{L}_{GR}$ is the reduced lagrangian depending on 
$g_{\text{tt}}(r)$, $g_{\text{rr}}(r)$ and $\rho(r)$ 
\ben \label{grlagr} \mathcal{L}_{GR}=\frac{1}{2}\left[\frac{1}{\sqrt{-g_{\text{rr}}}} 
\left(2\rho\rho'(\sqrt{g_{\text{tt}}})'+\sqrt{g_{\text{tt}}}(\rho')^2\right)+ 
\sqrt{-g_{\text{rr}}}\sqrt{g_{\text{tt}}} \right]. 
\een
\subsubsection{Electromagnetic Field Lagrangian}
Let's now consider the particle's rest frame so that the electromagnetic potential and the non 
vanishing components of the electromagnetic field tensor are 
\ben
\nonumber A_\mu(x)=(\varphi(r),{\bf 0}),\, \mathcal{F}_{rt}=-\mathcal{F}_{tr}=\varphi'(r). 
\een 
This gives a reduced action for the electromagnetic field 
\ben 
\nonumber \mathcal{A}_{EM}=-\frac{1}{16\pi }\int_{\mathbb{M}^{(1,3)}}d^4x\sqrt{|g^{(4)}|} 
\mathcal{F}_{\mu\nu}\mathcal{F}^{\mu\nu}=\int_{\mathbb{M}^{(1,1)}_{t,r}}dt\,dr\mathcal{L}_{EM}, 
\een 
where the reduced lagrangian is given by 
\ben \label{emlagr} 
\mathcal{L}_{EM}=\frac{1}{2}\frac{\rho(r)^2}{\sqrt{g_{\it{tt}}(r)} 
\sqrt{-g_{\text{rr}}(r)}}[\varphi'(r)]^2. 
\een
\subsection{Reduction of the Source's Action Integrals}
In order to obtain correct solution, the corresponding contributions of the point particle's 
self energy and the interaction with its own electromagnetic field have to be added to the total 
action. Later on it will be shown that the second, the so called electromagnetic self interaction, 
is well defined because of the regular behaviour of the electromagnetic field at the center of 
symmetry $r=0$. In other words our solution of this problem shows that there are no divergencies 
of the electromagnetic field at classical level. This is a consequence of the nontrivial geometry 
of spacetime. This may turn to play a role of a universal regularization factor because an infinite 
density of energy of any kind according to GR will cause similar deformations of spacetime. It turns 
out that this phenomena comes with another one which is disappearing of the horizons, i.e. no 
divergencies in the gravitational field either. These are the most important new effects under 
consideration in the present article.

The first of the above mentioned action integrals is well known and it can be written in the 
following form:
\ben 
\nonumber \mathcal{A}_{M_0}=-\int_{\mathbb{M}^{(1,4)}}\epsilon(r)\sqrt{|g^{(4)}|}d^4x=
-M_0\int_{\mathbb{L}^{(1)}}ds= \int_{\mathbb{M}^{(1,1)}_{t,r}}dt\, dr\mathcal{L}_{M_0}. 
\een 
Here $\mathbb{L}^{(1)}$ is the one-dimensional world-line of the particle and the second relation 
comes from the use of the three dimensional invariant Dirac delta function for the particle's mass 
density: 
\ben\label{barem} 
\epsilon(r)=M_0\delta_g(r)=\frac{M_0\delta(r)}{4\pi\sqrt{|g^{(3)}|}}=
\frac{M_0\delta(r)}{4\pi\sqrt{-g_{\text{rr}}}\rho(r)^2}. 
\een 
The mass $M_0$ in this formulation is the ``bare'' mass of the source corresponding to the energy of 
the static configuration. In other words it determines the energy needed to form the point source. 
Having this in mind the reduced lagrangian for the self energy is 
\ben\label{mlagr} 
\mathcal{L}_{M_0}=-M_0\sqrt{g_{\it{tt}}(r)}\delta(r). 
\een

The same consideration can be used to reduce the action describing the self interaction of the 
source 
\ben 
\nonumber A_{Q}=-\int_{\mathbb{M}^{(1,3)}}d^4x\sqrt{|g^{(4)}|}A_\mu(x)j^\mu(x)= 
-Q\int_{\mathbb{L}^{(1)}}A_\mu u^\mu ds=\int_{\mathbb{M}^{(1,1)}_{t,r}}dt\, dr\mathcal{L}_{Q}. 
\een 
This action integral and the action for the pure electromagnetic field are both manifestly covariant 
under diffeomorfisms of spacetime and $U(1)$ gauge group in electrodynamics. This is an important feature 
because it provides invariance of the solutions under such transformations with one restriction -- 
to consider only {\em regular} gauge transformations.

The charge current and charge density are given by 
\ben 
\nonumber j^\mu=q(r)u^\mu,\hskip .51truecm q(r)=Q\delta_g(r)=\frac{Q\delta(r)}{4\pi\sqrt{|g^{(3)}|}}=
\frac{Q\delta(r)}{4\pi\sqrt{-g_{\text{rr}}(r)}\rho(r)^2}, 
\een 
where the velocity $u^\mu$ is a unit length four-vector. If the particle's rest frame is chosen then 
$u^\mu=(\sqrt{g^{\text{tt}}},{\bf 0})$ so that the reduced lagrangian describing the self interaction is
\ben \label{qlagr} 
\mathcal{L}_{Q}=-Q\varphi(r)\delta(r). 
\een
\subsection{Field Equations and Quasinormal Coordinates}
To derive equations for the components of the metric tensor and the electromagnetic field of a point 
source the well known action principle will be used. This means that one has to sum equations 
(\ref{grlagr}), (\ref{emlagr}), (\ref{mlagr}) and (\ref{qlagr}) and find the variation of the reduced 
total action integral: $\mathcal{A}_{tot}=\int_{\mathbb{M}^{(1,1)}_{t,r}}dt\,dr\mathcal{L}_{tot}$ where 
\ben \label{totlagr} 
\mathcal{L}_{tot}:=\mathcal{L}_{GR}+\mathcal{L}_{EM}+\mathcal{L}_{M_0}+\mathcal{L}_{Q}. 
\een

Before doing this it is useful to transform the field variables $g_{\text{tt}}(r)$, $g_{\text{rr}}(r)$ 
and $\rho(r)$ to three new functions of the radial coordinate - $\varphi_1(r)$, $\varphi_2(r)$ and 
$\bar\varphi(r)$: 
\ben \label{mcompts}
\nonumber &&\sqrt{g}_{\text{tt}}=e^{\varphi_1},\, \sqrt{-g_{\text{rr}}}=
e^{2\varphi_2-\varphi_1-\bar{\varphi}},\,
\rho=\bar\rho e^{\varphi_2-\varphi_1}\\
&&\varphi_1=\ln(\sqrt{g_{\text{tt}}}),\, \varphi_2=\ln\left(\frac{\rho\sqrt{g_{\text{tt}}}}{\bar\rho}\right),\, 
\bar\varphi=\ln\left(\frac{\rho^2\sqrt{g_{\text{tt}}}}{\bar\rho^2\sqrt{-g_{\text{rr}}}}\right), 
\een 
which partly diagonalize the total lagrangian (\ref{totlagr}) and will be referred to as 
{\it{quasinormal coordinates}}. The symbol $\bar\rho$ denotes a normalization constant to be defined 
later from convenience reasons. The total lagrangian in terms of these new functions has the 
following structure: 
\ben \nonumber \mathcal{L}_{total}=\frac{1}{2}\Bigg[\bar\rho^2e^{\bar{\varphi}}\Big(-\varphi_1'^2+ 
\varphi_2'^2+e^{-2\varphi_1}\varphi'^2\Big)+e^{-\bar{\varphi}}e^{2\varphi_2}\Bigg]-\Big(M_0e^{\varphi_1}+
Q\varphi\Big) \delta(r). 
\een 
This lagrangian does not depend on derivatives of $\bar\varphi(r)$ so that the factors $e^{\pm\bar\varphi(r)}$ 
can be treated as Lagrange multipliers. This result comes from the above mentioned gauge freedom that was left.  
It can be fixed imposing a constraint in the lagrangian. According to the theory of constraints the field 
equation obtained from variation with respect to $\bar\varphi$ define a constraint on the solutions.

The full set of field equations are derived from variation of the total action with respect to all fields - 
$\varphi_1$, $\varphi_2$, $\varphi$, and $\bar\varphi$: 
\ben 
&&e^{-\bar{\varphi}}\frac{d}{dr}\left(e^{\bar{\varphi}}\frac{d\varphi_1}{dr}\right)= e^{-2\varphi_1}
\varphi'^2+\frac{M_0}{\bar\rho^2}e^{\varphi_1-\bar{\varphi}}\delta(r),
\nonumber\\
\nonumber &&e^{-\bar{\varphi}}\frac{d}{dr}\left(e^{\bar{\varphi}}\frac{d\varphi_2}{dr}\right)=
\frac{1}{\bar\rho^2}e^{2\varphi_1-2\bar{\varphi}},\hskip 3.3truecm\\
\nonumber &&e^{-\bar{\varphi}}\frac{d}{dr}\left(e^{\bar{\varphi}-\varphi_1}\frac{d\varphi}{dr}\right)=
\varphi_1'e^{-\varphi_1}\varphi'-\frac{Q}{\bar\rho^2}e^{\varphi_1-\bar{\varphi}}\delta(r),\\
\nonumber &&\bar\rho^2e^{\bar{\varphi}}\Big(-\half\varphi_1'^2+\half\varphi_2'^2+ \half e^{-2\varphi_1}\varphi'^2\Big)-
\half e^{2\varphi_2}e^{-\bar{\varphi}}\stackrel{w}{=}0, 
\een 
where the sign $\stackrel{w}{=}$ was used to denote the constraint.

Finally changing the electromagnetic field $\varphi(r)$ to a new variable $\varphi_3(r):\varphi_3'(r)=\varphi'(r)e^{-\varphi_1(r)}$ 
the full system of differential equations for the new fields in arbitrary gauge are 
\ben \label{rsystem} 
\nonumber
&&\bar\Delta_r\varphi_1=\varphi_3'^2+\frac{M_0}{\bar\rho^2}e^{\varphi_1-\bar\varphi}\delta(r),\\
&&\bar\Delta_r\varphi_2=\frac{1}{\bar\rho^2}e^{2\varphi_2-2\bar\varphi},\\
\nonumber
&&\bar\Delta_r\varphi_3=\varphi_1'\varphi_3'-\frac{Q}{\bar\rho^2}e^{\varphi_1-\bar\varphi}\delta(r),\\
\nonumber &&\half\bar\rho^2e^{\bar{\varphi}}\Big(-\varphi_1'^2+\varphi_2'^2+\varphi_3'^2\Big)-\half e^{2\varphi_2} 
e^{-\bar{\varphi}}\stackrel{w}{=}0. 
\een 
In these equations the differential operator 
$\bar\Delta_r:=e^{-\bar\varphi}{\frac{d}{dr}} \left( e^{\bar\varphi} {\frac{d}{dr}} \right)$ was introduced. 
This operator is connected to the radial part of the three dimensional Laplace--Beltrami operator: 
$\bar\Delta_r=-g_{\text{rr}}\Delta_r$, where $\Delta_r$ comes from the definition of the laplacian in 
curved 3D spacelike slices at a fixed time $t$: \\
$\Delta\!=\!-1/\sqrt{|{g}^{(3)}|}\partial_i \left(\sqrt{|{g}^{(3)}|}g^{ij}\partial_j\right)\!=\! -g^{ij}
\left(\partial^2_{ij}\!-\!\bar\Gamma_i\partial_j\right)\!= \!\Delta_r\!+\!1/\rho^2\Delta_{\theta\varphi}$. 
The four functions $\bar\Gamma_\mu$ appearing in this expression are not components of a vector field and their 
definition according to \cite{fock}, \cite{weinberg} is 
\ben 
\nonumber \bar\Gamma_\mu:=g_{\mu\alpha}g^{\beta\gamma}\Gamma^{\alpha}_{\beta\gamma}. 
\een 
One can fix their {\em local} values in an arbitrary way thus fixing the gauge of the problem, i.e. spacetime 
coordinates. In case of static spherically symmetric metric these functions are 
$\{\bar\Gamma_t,\bar\Gamma_r,\bar\Gamma_\theta,\bar\Gamma_\phi\}= \{0,\,\bar\varphi',\,\tan\theta,\,0\}$ which 
again shows that the variable $\bar\varphi(r)$ is the one fixing the radial gauge freedom. The global consequences 
of such gauge fixing need a very careful study because of the distributions appearing in the field equations and 
the possible divergencies they can provide.
\section{Gauge Fixing and First Integrals of the Field Equations}
In order to have regular field equations (\ref{rsystem}) one has to find a proper gauge condition. The proper 
choice of $\bar\varphi(r)$ must insure a finite multiplicative factor $e^{\varphi_1(r)-\bar\varphi(r)}$ of the 
Dirac delta function when $r$ goes to zero: 
\ben\label{gaugecond} 
\lim_{r\to 0}|\varphi_1(r)-\bar\varphi(r)|<\infty. 
\een 
This provides regular field equations at the place of the source $r=0$ \footnote{It is important to note that 
the center of spherical symmetry $r=0$ is mathematically excluded because of a coordinate singularity. In many problems 
it can be added back to the manifold by a trivial mapping but this problem requires a more sophisticated procedure 
due to the 3D invariant Dirac delta function placed there.}. The easiest way to find a proper gauge satisfying this 
is to solve the system (\ref{rsystem}) outside the source and take the limit $r\to 0$ with some assumptions for the 
asymptotic behaviour of $\bar\varphi(r)$.
\subsection{Solutions Outside the Source}
The first step in deriving solutions outside the source is to change the radial coordinate $r$ to a new variable 
$\sigma : d\sigma=e^{-\bar\varphi}dr$. This transforms $\bar\Delta_r$ to a new differential operator acting 
on functions of the new variable $\sigma$: $e^{-2\bar\varphi}\frac{d^2}{d\sigma^2}$ so that the field equations 
(\ref{rsystem}) outside the source in terms of $\sigma$ read 
\ben\label{sigmasystem} \nonumber
&&\frac{d^2\varphi_1}{d\sigma^2}=\left(\frac{d\varphi_3}{d\sigma}\right)^2,\\
&&\frac{d^2\varphi_2}{d\sigma^2}=\frac{1}{\bar\rho^2}e^{2\varphi_2},\\
\nonumber &&\frac{d^2\varphi_3}{d\sigma^2}=\left(\frac{d\varphi_1}{d\sigma}\right)
\left(\frac{d\varphi_3}{d\sigma}\right),\\
\nonumber &&\half\bar\rho^2\left(-\left(\frac{d\varphi_1}{d\sigma}\right)^2+\left(\frac{d\varphi_2}{d\sigma}\right)^2+ 
\left(\frac{d\varphi_3}{d\sigma}\right)^2\right)- \half e^{2\varphi_2}\stackrel{w}{=}0. 
\een

Next it is useful to define variables which can be interpreted as {\it{canonical energy}} for the corresponding 
field such that 
\ben \label{epsilons} \varepsilon_1:=-\half\bar\rho^2\left(\frac{d\varphi_1}{d\sigma}\right)^2,\, 
\varepsilon_2:=\half\left(\bar\rho^2\left(\frac{d\varphi_2}{d\sigma}\right)^2-e^{2\varphi_2}\right),\, \varepsilon_3:=
\half\bar\rho^2\left(\frac{d\varphi_3}{d\sigma}\right)^2. 
\een A proper substitution of these in (\ref{sigmasystem}) combines $\varepsilon_1$, $\varepsilon_2$ and $\varepsilon_3$ 
into two first integrals: $\varepsilon_2=\epsilon_2=\text{const}$ and $\varepsilon_1+\varepsilon_3=-\epsilon_2$. 
As usual these two first integrals are used to divide (\ref{sigmasystem}) to two separate equations which in the 
current case are exactly the same while the equation for $\varphi_3$ still depends on $\varphi_1$: 
\ben 
\nonumber \frac{d^2}{d\sigma^2}e^{-\varphi_{1,2}}-\frac{2\epsilon_2}{\bar\rho^2}e^{-\varphi_{1,2}}=0,\,\,\, 
\left(\frac{d\varphi_3}{d\sigma}\right)^2=\left(\frac{d\varphi_1}{d\sigma}\right)^2-\frac{2\epsilon_2}{\bar\rho^2}. 
\een 
These are the final field equations for the charged point particle outside the source. Depending on the sign of 
$\epsilon_2$ this system has three different types of solutions. Later on it will become clear that this sign corresponds 
to charges higher, equal or lower than the mass in terms of the usual parameters used for description of the solution. 
These three solutions are:

I. In case $\epsilon_2>0$: 
\ben \nonumber
&&\varphi_1(\sigma)=C_1-\ln\left(\sinh(\frac{\sqrt{2\epsilon_2}}{\bar\rho}\sigma+C_2)\right),\\
\nonumber
&&\varphi_2(\sigma)=C_3-\ln\left(\sinh(\frac{\sqrt{2\epsilon_2}}{\bar\rho}\sigma+C_4)\right),\\
\nonumber &&\varphi(\sigma)=\mp\frac{e^{C_1}}{\tanh(\frac{\sqrt{2\epsilon_2}}{\bar\rho}\sigma+C_2)}+C_5. 
\een

II. In case $\epsilon_2=0$: 
\ben \nonumber
&&\varphi_1(\sigma)=-\ln\left(C_1\sigma+C_2\right),\\
\nonumber
&&\varphi_2(\sigma)=-\ln\left(C_3\sigma+C_4\right),\\
\nonumber &&\varphi(\sigma)=\mp\frac{C_1}{C_1\sigma+C_2}+C_5. 
\een

III. In case $\epsilon_2<0$: 
\ben \nonumber
&&\varphi_1(\sigma)=C_1-\ln\left(\sin(\frac{\sqrt{-2\epsilon_2}}{\bar\rho}\sigma+C_2)\right),\\
\nonumber
&&\varphi_2(\sigma)=C_3-\ln\left(\sin(\frac{\sqrt{-2\epsilon_2}}{\bar\rho}\sigma+C_4)\right),\\
\nonumber &&\varphi(\sigma)=\mp\frac{e^{C_1}}{\tan(\frac{\sqrt{-2\epsilon_2}}{\bar\rho}\sigma+C_2)}+C_5. 
\een

\subsection{Basic Regular Gauge for the Point Particle}
After solving (\ref{rsystem}) outside the source one has to find $\bar\varphi(r)$ satisfying (\ref{gaugecond}). 
A large class of gauge conditions is described by lagrange multipliers $e^{\pm\bar\varphi(r)}$ with the following 
behaviour when $r\to 0$: $e^{-\bar\varphi(r)}\sim k r^n$. Considering this gauge class which is general enough 
there are two possibilities for the power $n$ of $r$ in the asymptotic behaviour

I) If $n\ne -1$ is chosen it implies $\sigma=\frac{k}{n+1}r^{n+1}$ so that $|\varphi_1(r)-\bar\varphi(r)|$ is 
\ben \nonumber &&\Big|C_1+\ln{k}+\ln\frac{r^n}{\sinh(\frac{k\sqrt{-2\epsilon_2}}{(n+1)\bar\rho}r^{n+1}+
C_2)}\Big|,\\
\nonumber
&&\Big|\ln{k}+\ln\frac{r^n}{\frac{kC_1}{n+1}r^{n+1}+C_2}\Big|,\\
\nonumber &&\Big|C_1+\ln{k}+\ln\frac{r^n}{\sin(\frac{k\sqrt{-2\epsilon_2}}{(n+1)\bar\rho}r^{n+1}+ C_2)}\Big|, 
\een 
for each of $\epsilon_2>0$, $\epsilon_2=0$, $\epsilon_2<0$ respectively. This means that the only choice for 
$n$ that satisfies the gauge condition (\ref{gaugecond}) independently from the integrating constants is $n=0$.

II) The case $n=-1$ corresponds to $\sigma=k\ln{r}$ which always gives divergent value of 
$|\varphi_1(r)-\bar\varphi(r)|$ when $r$ goes to zero.

Finally the only choice that keeps the fields regular at $r=0$ is $e^{-\bar\varphi(r)}\sim const$ for 
$r\rightarrow 0$. This defines a whole class of possible gauges which will be referred as \textit{regular gauges} 
for the point particle source. The easiest way to choose one from those regular gauges is to impose the condition 
$e^{-\bar\varphi(r)}=1$ or simply 
\ben\label{brg} \bar\varphi(r)=0. 
\een 
This choice is the so called \textit{Basic Regular Gauge (BRG)} \cite{fiziev}. It will be used from now on for 
solving the Einstein-Maxwell point particle problem.

\subsection{Field Equations in BRG and First Integrals}
The next step after fixing the gauge freedom according to (\ref{brg}) is to solve the field equations 
(\ref{rsystem}) in BRG. The latter read 
\ben \label{fsystem} \nonumber
&&\varphi_1''=\varphi_3'^2+\frac{M_0}{\bar\rho^2}e^{\varphi_1}\delta(r),\\
&&\varphi_2''=\frac{1}{\bar\rho^2}e^{2\varphi_2}\\
\nonumber
&&\varphi_3''=\varphi_1'\varphi_3'-\frac{Q}{\bar\rho^2}e^{\varphi_1}\delta(r),\\
\nonumber &&\half\bar\rho^2\Big(-\varphi_1'^2+\varphi_2'^2+\varphi_3'^2\Big)-\half e^{2\varphi_2}\stackrel{w}{=}0. 
\een

In order to divide these equations into independent parts conserved quantities have to be extracted. These are first 
integrals for this system and they can be easily derived from combining (\ref{epsilons}) where the choice of BRG gives 
$\sigma=r$. In terms of $\varepsilon_{1,2,3}$ (\ref{fsystem}) becomes 
\ben \label{help1} \nonumber &&\varphi_2''-\varphi_1''-\left(\varphi_2'^2-\varphi_1'^2\right)=-\frac{2}{\bar\rho^2} 
\left(\varepsilon_1+\varepsilon_2+\varepsilon_3\right)-
\frac{M_0}{\bar\rho^2}e^{\varphi_1}\delta(r),\\
&&\varepsilon_2'=0,\\
\nonumber &&\left(\varepsilon_1+\varepsilon_3\right)'=-\frac{1}{\bar\rho^2}\left(M_0\varphi_1'+
Q\varphi_3'\right)e^{\varphi_1}\delta(r),\\
\nonumber &&\varepsilon_1+\varepsilon_2+\varepsilon_3\stackrel{w}{=}0. 
\een 
Because of the Dirac delta functions in the latter it is useful to introduce a new quantity $\Delta(r):=\theta(r)-1$ 
which has the following properties \ben \label{newheav} \nonumber &1)&\, \, \, \Delta^{(n)}(r)=\delta^{(n-1)}(r),\,\,
\text{supp}\left(\Delta^{(n)}(r)\right)={0},\quad n\ge 1\\
&2)&\, \, \, \left(\Delta(r)\right)^n=(-1)^{n-1}\Delta(r),\quad n\ge 1. 
\een 
This is a function which is zero for all values of $r\ge 0$ and $-1$ for $r<0$.

With the use of (\ref{newheav}) and the definition of quasinormal coordinates (\ref{mcompts}) the equations (\ref{help1}) 
transform to 
\ben \nonumber &&\left(\frac{\rho'(r)}{\sqrt{g_{\text{rr}}(r)g_{\text{tt}}(r)}}+
\frac{M_0}{\bar\rho}e^{-\varphi_2(0)}\Delta(r)\right)'=-\frac{2\varepsilon_{\text{tot}}}{\bar\rho}e^{(-\varphi_2-\varphi_1)},\\
\nonumber &&\varepsilon_{\text{tot}}+\frac{1}{\bar\rho^2}\left(M_0\varphi_1'(0)+
Q\varphi_3'(0)\right)e^{\varphi_1(0)}\Delta(r)=E_{\text{tot}},\\
\nonumber
&&\varepsilon_{\text{tot}}:=\varepsilon_1+\varepsilon_2+\varepsilon_3\stackrel{w}{=}0,\\
\nonumber &&\varepsilon_2=\epsilon_2, 
\een 
where $\epsilon_2$ and $E_\text{tot}$ are constants. A combination of the second and third of these equations implies 
\ben\nonumber E_{\text{tot}}\stackrel{w}{=}\frac{1}{\bar\rho^2}\left(M_0\varphi_1'(0)+ Q\varphi_3'(0)\right)e^{\varphi_1(0)}
\Delta(r) 
\een 
so that the choice of a domain for $r$ such that $r\in[0,\infty)$ determines only one possible value of $E_\text{tot}=0$. 
With this value and the fact that $\Delta(r)=0,\, \forall r\in[0,\infty)$\footnote{The function is zero but its 
derivative is the Dirac delta function} there are three first integrals for the system of field equations 
\ben \label{firstints} \frac{\rho'(r)}{\sqrt{g_{\text{rr}}(r)g_{\text{tt}}(r)}}\stackrel{w}{=}C=\text{const},\,\, 
\varepsilon_2=\epsilon_2=\text{const},\,\, \varepsilon_1+\epsilon_2+\varepsilon_3\stackrel{w}{=}0. 
\een 
Finally using the second and third of these and the definition (\ref{epsilons}) the system transforms to a very 
``easy to integrate'' one: \ben \label{fnalsystem} \nonumber &&\frac{d^2}{dr^2}e^{-\varphi_1}-\frac{2\epsilon_2}
{\bar\rho^2}e^{-\varphi_1}
-\frac{M_0}{\bar\rho^2}\delta(r)=0,\\
&&\frac{d^2}{dr^2}e^{-\varphi_2}-\frac{2\epsilon_2}{\bar\rho^2}e^{-\varphi_2}=0,\\
\nonumber &&\varphi_3'^2=\varphi_1'^2-\frac{2\epsilon_2}{\bar\rho^2}. 
\een

\section{Solutions for the Gravitational Field}
The solutions of the field equations corresponding to point particle are these solutions of (\ref{fnalsystem}) that 
satisfy the proper asymptotic conditions, i.e. the physics of the problem will determine the integration constants. 
It was shown that the general solution of the field equations (\ref{fnalsystem}) outside the source depends on the 
sign of $\epsilon_2$ so that there are three cases that should be considered:

I) Positive values, $\epsilon_2>0$.

\noindent Later on it will be shown that these values of $\epsilon_2$ correspond to charges smaller than the mass of 
the point source. The solutions of (\ref{fnalsystem}) for the gravitational field in this case are 
\ben \nonumber &&e^{-\varphi_1(r)}=C_1\sinh{\left(\Omega\left(r_1-r\right)\right)}+\frac{M_0}{\bar\rho^2\Omega}\sinh
\left({\Omega r}\right)\Delta(r),\\
\nonumber &&e^{-\varphi_2(r)}=C_2\sinh{\left(\Omega\left(r_2-r\right)\right)}, 
\een 
where $\Omega:=\sqrt{\frac{2\epsilon_2}{\bar\rho^2}}$ and $C_1$, $C_2$, $r_1$ and $r_2$ are constants coming from 
integration of the equations.

II) Zero value, $\epsilon_2=0$.

\noindent This corresponds to equal mass and charge and the solutions are 
\ben \nonumber
&&e^{-\varphi_1(r)}=C_1\left(r_1-r\right)+\frac{M_0}{\bar\rho^2}r\Delta(r),\\
\nonumber &&e^{-\varphi_2(r)}=C_2\left(r_2-r\right). 
\een

III) Negative values, $\epsilon_2<0$.

\noindent This is the most physically significant case because it corresponds to charges greater than the mass of the 
point particle. This is just the observed relation for real world particles. Here the solutions for the gravitational 
field are 
\ben \nonumber &&e^{-\varphi_1(r)}=C_1\sin{\left[\Omega\left(r_1-r\right)\right]}+\frac{M_0}{\bar\rho^2\Omega}
\sin\left({\Omega r}\right)\Delta(r),\\
\nonumber &&e^{-\varphi_2(r)}=C_2\sin{\left[\Omega\left(r_2-r\right)\right]}, 
\een 
where $\Omega:=\sqrt{-\frac{2\epsilon_2}{\bar\rho^2}}$.

The integration in each of these cases is not straightforward because of the delta function. The way to solve 
(\ref{fnalsystem}) is to fourier transform the equation with a delta function and find a proper solution for 
each case. The final solution is derived by adding this result to the solution of the homogenous equation.

\subsection{Asymptotic Behaviour}
As was mentioned above the integration constants should be obtained from some physical conditions on the general 
solution. The first two such conditions are asymptotic flatness of the metric and coincidence with the vacuum 
solution in terms of $t,\rho,\theta,\phi$ outside the source, i.e. coincidence with the solution written in the 
usual Reissner-Nordstr$\ddot{\text{o}}$m form. These are quite common while the third condition is less trivial 
and was recently discussed in \cite{fiziev}.

The asymptotic behaviour of the solutions for the gravitational field is defined via the limit $\rho(r)\to\infty$ 
so that it should be obtained for each case individually.

Let's first consider the case $\epsilon_2>0$. Having in mind the expression for $\rho(r)$ in terms of $\varphi_1$ 
and $\varphi_2$ (\ref{mcompts}) one has 
\ben \nonumber \rho(r)=\frac{C_1\sinh{\left[\Omega\left(r_1-r\right)\right]}+ \frac{M_0}{\bar\rho^2\Omega}
\sinh\left({\Omega r}\right)\Delta(r)}{C_2\sinh{\left[\Omega\left(r_2-r\right)\right]}} \een and it is obvious 
that $\rho(r)\to \infty$ when $r\to r_2=:r_\infty$.

In the second case $\epsilon_2=0$ one arrives at 
\ben \nonumber \rho(r)=\frac{C_1\left(r_1-r\right)+\frac{M_0}{\bar\rho^2}r\Delta(r)} {C_2\left(r_2-r\right)} 
\een 
and again $\rho(r)\to \infty$ when $r=\to r_2=:r_\infty$.

Finally when $\epsilon_2<0$ 
\ben \nonumber \rho(r)=\frac{C_1\sin{\left[\Omega\left(r_1-r\right)\right]}+ \frac{M_0}
{\bar\rho^2\Omega}\sin\left({\Omega r}\right)\Delta(r)}{C_2\sin{\left[\Omega\left(r_2-r\right)\right]}} 
\een so that $\rho(r)\to \infty$ when $r\to r_2-\frac{k\pi}{\Omega}$. In order to have divergence-free solutions one 
has to choose the value of $k$ such that 
$r\to\frac{\pi}{\Omega}\left(\frac{r_2\Omega}{\pi}- \left[\frac{r_2\Omega}{\pi}\right]\right)=:r_\infty$, where 
$\left[\frac{p}{q}\right]$ means the integer part of $\frac{p}{q}$. This is so because using this choice $r$ belongs 
to the interval $\left(0,r_\infty\right]$ and there are no divergent points in it, i.e. $\Omega(r_2-r)\in[0,\pi)$ when 
$r\in\left(0,r_\infty\right]$.

These results are quite significant because they mean that the physical infinity for the problem is at a finite 
value of the radial coordinate. In other words the regular gauge condition determines a finite interval 
$(0,r_\infty]$ for the radial coordinate in which the {\em{regular}} point-like solutions live. Later on it will 
become clear that this restriction for the domain leads to a finite value of $\rho(r)$ when the center of symmetry 
$r=0$ is added to the manifold. This means nothing else but finite value of the luminosity of the point source. The 
reason to refer to $\rho(r)$ as a \textit{luminosity variable} is that it determines the luminosity via the expression 
$L=\frac{I}{4\pi\rho(r)^2}$, where $I$ is the intensity of the source.

At this point it is crucial to note that the finite domain for $r$ is where arbitrary {\em regular} coordinate 
transformations (gauges) are allowed. This is due to the regularity conditions that each gauge theory is forced 
to satisfy, i.e. regularity of the gauge transformation in the domain where the physical fields live. Thus up 
to now only the domain of the gravitational gauge invariant fields, the metric components, was defined. This 
domain is easy to be mapped on a new one, where $r$ varies from $0$ to $\infty$.  This can be done by using a proper 
{\em regular} transformation.

After defining $r_\infty$ as the end of the {\em regular} physical domain for $r$ one has to impose flatness of the 
solution in the limit $\rho(r)\to\infty$ or equivalently $r=r_\infty$. This means that at $r=r_\infty$ the metric 
components have to coincide with the flat ones: $g_{\text{tt}}(r_\infty)=1$ and $g_{\rho\rho}(r_\infty)=-1$, where 
$g_{\rho\rho}\rho'^2=g_{\text{rr}}$. Now using these conditions and (\ref{firstints}) one obtains 
\ben \nonumber C=\frac{\rho'(r)}{\sqrt{-g_{\text{rr}}(r)g_{\text{tt}}(r)}}=1. 
\een 
Finally expressing $g_{\text{tt}}(r)$ and the above first integral in terms of $\varphi_1(r)$ and $\varphi_2(r)$ the 
asymptotic flatness gives the following two equations 
\ben \nonumber e^{-\varphi_1(r_\infty)}=1,\,\,\bar\rho\left[\left(e^{-\varphi_1(r)}\right)'e^{-\varphi_2(r)}- 
\left(e^{-\varphi_2(r)}\right)'e^{-\varphi_1(r)}\right]\Bigg|_{r\to r_\infty}=1. 
\een 
Solving these determines the constants $C_1$ and $C_2$ for each of the three cases and the solutions of 
(\ref{fnalsystem}) satisfying the above mentioned asymptotic conditions are:

1) $\epsilon>0$ 
\ben 
\nonumber &&e^{-\varphi_1(r)}=\frac{\sinh{\left[\Omega\left(r_1-r\right)\right]}} {\sinh{\left[\Omega
\left(r_1-r_\infty\right)\right]}}+
\frac{M_0}{\bar\rho^2\Omega}\sinh\left(\Omega r\right)\Delta(r),\\
\nonumber &&e^{-\varphi_2(r)}=\frac{1}{\bar\rho\Omega}\sinh{\left[\Omega\left(r_\infty-r\right) \right]}. 
\een

2) $\epsilon=0$ 
\ben 
\nonumber &&e^{-\varphi_1(r)}=\frac{r_1-r}{r_1-r_\infty}+
\frac{M_0}{\bar\rho^2}r\Delta(r),\\
\nonumber &&e^{-\varphi_2(r)}=\frac{r_\infty-r}{\bar\rho}. 
\een

3) $\epsilon<0$ 
\ben 
\nonumber &&e^{-\varphi_1(r)}=\frac{\sin{\left[\Omega\left(r_1-r\right)\right]}} 
{\sin{\left[\Omega\left(r_1-r_\infty\right)\right]}}+
\frac{M_0}{\bar\rho^2\Omega}\sin\left(\Omega r\right)\Delta(r),\\
\nonumber &&e^{-\varphi_2(r)}=\frac{1}{\bar\rho\Omega}\sin{\left[\Omega\left(r_\infty-r\right) \right]}. 
\een
\subsection{Coincidence with the Vacuum Solution Outside the Source}
The next step after determining $C_1$ and $C_2$ from asymptotic flatness is to impose that outside the source 
the solution for $g_{\text{tt}}$ is the same in terms of $(t,\rho,\theta,\phi)$ as the well known form of the vacuum 
Reissner-Nordstr$\ddot{\text{o}}$m solution. One should note here that the coordinate transformation from the latter to 
the current solution is defined only in the domain $r\in(0,r_\infty]$ which maps to $\rho\in(\rho_0,\infty)$ with 
a non-zero value of $\rho_0\equiv\rho(0)$. This means that the current solution is regular for $r=0$ and later on 
will be shown that it has no singularities in the region $\rho\in(\rho_0,\infty)$.

1) Consider the case $\epsilon_2>0$, so that $\Omega$ is defined as $\Omega=\sqrt{\frac{2\epsilon_2}{\bar\rho^2}}$. 
The solutions for $g_{\text{tt}}(r)$ and $\rho(r)$ outside the source are 
\ben\nonumber 
g_{\text{tt}}(r)=\frac{\sinh^2\big[\Omega(r_1-r_\infty)\big]}{\sinh^2\big[\Omega (r_1-r)\big]},\,\, 
\rho(r)=\bar\rho\frac{\sinh\big[\Omega (r_1-r)\big]} {\sinh\big[\Omega(r_1-r_\infty)\big]}\frac{\Omega \bar\rho}
{\sinh\big[\Omega (r_\infty-r) \big]}. 
\een 
After simple algebraic calculations the following relation can be derived 
\ben\label{rnsol} g_{\text{tt}}(r)=1-2\frac{\Omega\bar\rho^2\coth\big[\Omega(r_1-r_\infty)\big]}{\rho(r)}+ 
\frac{\left(\Omega\bar\rho^2\sinh^{-1}\big[\Omega(r_1-r_\infty)\big]\right)^2}{\rho(r)^2}. 
\een 
Comparing this result to the well known expression $g_{\text{tt}}=1-\frac{2M}{\rho}+\frac{Q^2}{\rho^2}$ one ends up 
with the following system for the unknown parameters $\Omega$, $r_1$, $r_1-r_\infty$ and $\bar\rho$ 
\ben 
\nonumber
&&\Omega\bar\rho^2\coth\big[\Omega(r_1-r_\infty)\big]=M,\\
\nonumber &&\frac{\Omega\bar\rho^2}{\sinh\big[\Omega(r_1-r_\infty)\big]}=|Q|. 
\een 
Here $M$ is the \textit{Keplerian mass} \cite{landau} of the source, which in general is different from the ``bare'' mass $M_0$, 
and $Q$ is the electric charge. The solutions of these two equations for $\Omega$ and $r_1-r_\infty$ are 
\ben\nonumber \Omega=\frac{M}{\bar\rho^2}\sqrt{1-\eta^2},\,\,\coth\big[\Omega(r_1-r_\infty)\big]=\frac{1}{\sqrt{1-\eta^2}}, 
\een 
where $\eta$ is a dimensionless parameter such that \ben\label{etadef} \eta:=\frac{|Q|}{M}. \een Because of 
$\epsilon_2>0$ this solution restricts $\eta$ to values smaller than $1$ which is simply the case of $Q<M$.

With these parameters set the solutions for the gravitational field corresponding to $\eta<1$ are 
\ben \nonumber 
e^{-\varphi_1(r)}&=&\cosh\left[\Omega\left(r_\infty-r\right)\right]+
\frac{1}{\sqrt{1-\eta^2}}\sinh\left[\Omega\left(r_\infty-r\right)\right]+\\
\nonumber
&+&\frac{M_0}{M\sqrt{1-\eta^2}}\sinh\left(\Omega r\right)\Delta(r),\\
\nonumber e^{-\varphi_2(r)}&=&\frac{\frac{\bar\rho}{M}}{\sqrt{1-\eta^2}}
\sinh{\left[\Omega\left(r_\infty-r\right)\right]},\\
\nonumber \Omega&:=&\frac{M}{\bar\rho^2}\sqrt{1-\eta^2}. 
\een

2) In the case $\epsilon_2=0$ there is an expression for $g_{\text{tt}}(\rho)$ similar to (\ref{rnsol})is 
\ben g_{\text{tt}}(r)=1-2\frac{\frac{\bar\rho^2}{r_1-r_\infty}}{\rho(r)}+ \frac{\frac{\bar\rho^4}{(r_1-r_\infty)^2}}
{\rho(r)^2}, 
\een 
which corresponds to $M=Q$ so that $r_1-r_\infty=\frac{\bar\rho^2}{M}$. According to this the solutions of the field 
equations are 
\ben\nonumber
&&e^{-\varphi_1(r)}=1+\frac{M}{\bar\rho^2}\left[r_\infty-r+\frac{M_0}{M}r\Delta(r)\right],\\
\nonumber &&e^{-\varphi_2(r)}=\frac{r_\infty-r}{\bar\rho}. 
\een

3) The case $\epsilon_2<0$ is very similar to 1). The difference is in $\Omega=\sqrt{-\frac{2\epsilon}{\bar\rho^2}}$ 
and the equations for the integrating constants are 
\ben \nonumber
&&\Omega\bar\rho^2\cot\big[\Omega(r_1-r_\infty)\big]=M,\\
\nonumber &&\frac{\Omega\bar\rho^2}{\sin\big[\Omega(r_1-r_\infty)\big]}=|Q|. 
\een 
The solutions of these two equations are 
\ben \Omega=\frac{M}{\bar\rho^2}\sqrt{\eta^2-1},\,\,\cot\big[\Omega(r_1-r_\infty)\big]=\frac{1}{\sqrt{\eta^2-1}}, 
\een 
where the dimensionless parameter $\eta$ (\ref{etadef}) is greater than one corresponding to $\epsilon_2<0$.

The final expression for the solutions is 
\ben \nonumber e^{-\varphi_1(r)}&=&\cos\left[\Omega\left(r_\infty-r\right)\right]+
\frac{1}{\sqrt{\eta^2-1}}\sin\left[\Omega\left(r_\infty-r\right)\right]+\\
\nonumber
&+&\frac{M_0}{M\sqrt{\eta^2-1}}\sin\left(\Omega r\right)\Delta(r),\\
\nonumber e^{-\varphi_2(r)}&=&\frac{\frac{\bar\rho}{M}}{\sqrt{\eta^2-1}}
\sin{\left[\Omega\left(r_\infty-r\right)\right]},\\
\nonumber \Omega&:=&\frac{M}{\bar\rho^2}\sqrt{\eta^2-1}. 
\een

\subsection{Gravitational Mass Defect of a Point Source}
It is a well known fact that there is a difference between the mass in the Einstein equations and the so called 
``bare'' mass which is the invariant quantity defined as an integral of the mass density. The inequality of these 
two masses defines a new free parameter in the solution which will be considered in this subsection.

First of all in case of spherical symmetry one has the following expression for the bare mass 
\ben\label{spmass} M_0=4\pi\int \sqrt{-g_{\text{rr}}(r)}\,\rho^2(r)\,\epsilon(r)dr, 
\een 
where $\epsilon(r)$ is the invariant density distribution defined in (\ref{barem}). This is the quantity describing 
the energy of the particles configuration as discussed above.

Besides the bare mass one can consider the mass in Einstein equations which is the so called \textit{Keplerian mass} 
of the source and is the one that an observer at infinity will measure witnessing the motion of a test particle. In 
case of a point source the Keplerian mass is 
\ben \label{kmass} M=4\pi\int\rho^\prime(r)\rho^2(r)\epsilon(r)dr=M_0\int\sqrt{g_{\text{tt}}(r)}\delta(r)dr= 
M_0\sqrt{g_{\text{tt}}(0)}. 
\een

The dimensionless quantity $\varrho:=\sqrt{g_{\text{tt}}(0)}$ takes its values in the interval $[0,1]$ and will be 
referred to as ``gravitational mass defect ratio''. Later on it will be shown that the values $\varrho=0,1$ are not 
physical, i.e. they give divergencies in the metric components.

Now using the definition of $\varrho$ and performing a rational change of the radial coordinate $r\to\zeta$ which 
maps the finite domain $r\in(0,r_\infty]$ to $\zeta\in(0,\infty)$: 
\ben \label{ratch} \zeta:=\frac{r/r_\infty}{1-r/r_\infty},\,\, r=r_\infty\frac{\zeta}{\zeta+1} 
\een 
gives the following solutions depending on sign($\eta$):

1) $\eta<1$ 
\ben \label{etal} \nonumber &&e^{-\varphi_1(\zeta)}=\cosh\left(\frac{\Delta\Phi^-}{\zeta+1}\right)+ 
\frac{\sinh\left(\frac{\Delta\Phi^-}{\zeta+1}\right)}{\sqrt{1-\eta^2}}+ 
\frac{\sinh\left(\frac{\Delta\Phi^-\zeta}{\zeta+1}\right)}{\varrho\sqrt{1-\eta^2}}
\Delta(\zeta),\\
&&e^{-\varphi_2(\zeta)}=\frac{1}{\xi\sqrt{1-\eta^2}}
\sinh{\left(\frac{\Delta\Phi^-}{\zeta+1}\right),}\\
\nonumber &&\Delta\Phi^-:=\text{arcsinh}{\frac{\sqrt{1-\eta^2}}{\eta\varrho}}- \text{arcsinh}{\frac{\sqrt{1-\eta^2}}{\eta}}. 
\een

2) $\eta=1$ 
\ben \label{etae} \nonumber &&e^{-\varphi_1(\zeta)}=1+\frac{1-\varrho}{\varrho}\left(\frac{1}{\zeta+1}+
\frac{1}{\varrho}\frac{\zeta}{\zeta+1}
\Delta(\zeta)\right),\\
&&e^{-\varphi_2(\zeta)}=\frac{1-\varrho}{\xi\varrho}\frac{1}{\zeta+1}. 
\een

3) $\eta>1$ \ben \label{etag} \nonumber &&e^{-\varphi_1(\zeta)}=\cos\left(\frac{\Delta\Phi^+}{\zeta+1}\right)+ 
\frac{\sin\left(\frac{\Delta\Phi^+}{\zeta+1}\right)}{\sqrt{\eta^2-1}}+
\frac{\sin\left(\Delta\Phi^+\frac{\zeta}{\zeta+1}\right)}{\varrho\sqrt{\eta^2-1}}\Delta(\zeta),\\
&&e^{-\varphi_2(\zeta)}=\frac{1}{\xi\sqrt{\eta^2-1}}
\sin{\left(\frac{\Delta\Phi^+}{\zeta+1}\right)},\\
\nonumber &&\Delta\Phi^+:=\text{arcsin}{\frac{\sqrt{\eta^2-1}}{\eta\varrho}}-
\text{arcsin}{\frac{\sqrt{\eta^2-1}}{\eta}},\\
&&\nonumber \eta\le\frac{1}{\sqrt{1-\varrho^2}}, 
\een 
where $\xi:=\frac{M}{\bar\rho}$ in all cases.

The consideration for $\eta>1$ is the most interesting one because it corresponds to charges higher than the 
mass of the point particle. This is the most common case in real world particle physics. In this solution 
there is a limitation for the charge of the point particle given by the last inequality. It comes from the 
relation $e^{\varphi(0)}=1/\varrho$ between $\Omega r_\infty$ and the dimensionless parameters: 
\ben \label{condtrig} \sin\left(\arcsin{\frac{\sqrt{\eta^2-1}}{\eta}}+\Omega r_\infty\right)= \frac{\sqrt{\eta^2-1}}{\eta\varrho}. 
\een 
This equation has real solutions for $\Omega r_\infty$ only if the right hand side is smaller than $1$ so that 
$\sqrt{\eta^2-1}\le{\eta\varrho}$. It is interesting to note here that this can be considered as a limit for the 
charge of the point particle: \ben \nonumber |Q|\le\frac{M}{\sqrt{1-\varrho^2}}. \een This limiting value depends 
on the gravitational mass defect $\varrho$ which is very small for real particles because of their weak 
gravitational field so that their charges can be big enough.

The luminosity variable in all these cases does not depend on $\bar\rho$ and it is useful to define a dimensionless 
quantity $z(\zeta)$: 
\ben\nonumber z(\zeta):=\frac{\rho(\zeta)}{M}\equiv\frac{e^{-\phi_1(\zeta)}}{\xi e^{-\phi_2(\zeta)}}, 
\een 
which is just $\rho(\zeta)$ measured in units of $M$. This function outside the source is given by:
\begin{itemize}
\item{$\eta<1$}, $z(\zeta)=1+\sqrt{1-\eta^2}\coth\left(\frac{\Delta\Phi^-}{\zeta+1}\right)$,
\item{$\eta=1$}, $z(\zeta)=1+\frac{\varrho}{1-\varrho}\left(\zeta+1\right)$,
\item{$\eta>1$}, $z(\zeta)=1+\sqrt{\eta^2-1}\cot\left(\frac{\Delta\Phi^+}{\zeta+1}\right)$.
\end{itemize}
The value of $z(0)$ is the same for each of these three cases: 
\ben \label{zzero} z(0)=\frac{\eta^2}{1-\sqrt{1-\eta^2(1-\varrho^2)}}, 
\een 
where in case 2) $z(0)$ is given by the simple substitution $\eta=1$ in the last expression.

In case of particles, with $\eta>1$ a limiting value of the gravitational mass defect ratio exists: 
\ben\nonumber \varrho\ge\varrho_{\text{lim}}=\sqrt{1-\frac{1}{\eta^2}}, 
\een 
which can be derived in the same way as the charge limit discussed above. The physical relevance of this limiting 
value becomes clear after its substitution in (\ref{zzero}). It gives a limit for the initial value of the luminosity 
variable which is 
\ben \rho_{0}^{\text{lim}}=\frac{Q^2}{M}. 
\een 
This is exactly the classical radius of the particle derived from elementary energy conservation assumptions - 
$\rho_{\text{cl}}=Q^2/M$. Within the current framework this radius corresponds to the minimal value 
$\rho_0\ge\rho_0^{\text{lim}}$ of the luminosity variable due to the gravitational field produced by the point source. 
In the next section it will be shown that this provides a cut for the corresponding electromagnetic potential and 
thus regular behaviour at the point $r=0$. This leads to a cut of the electromagnetic field if the motion of a charged 
test particle in the presence of point-like source is considered. 

Another interesting consequence when $\eta<1$, is that the cut in $\rho(r)$ is always outside the event horizon appearing 
in the vacuum solution without any restrictions for the gravitational mass defect ratio $\varrho$: 
\ben \nonumber z(0)>1+\sqrt{1-\eta^2}.
\een 
This corresponds to disappearing of the coordinate singularities which arise in the physical domain in the classical 
Reissner-Nordstr$\ddot{\text{o}}$m form of the vacuum solution. For our solution the horizon and all phenomena related 
with it are in the non-physical domain of the variables. It is important at this point to note that the solution 
derived here and the Reissner-Nordstr$\ddot{\text{o}}$m one are the same outside the source in terms of $t,\rho,\theta$ 
and $\phi$ but there is a significant difference between them because of the above mentioned cut on $\rho(r)$. In case 
of the classical solution there is a $\delta(\rho)/\rho$ pathological divergence at the point source placed in $\rho=0$ 
while the current solution has only $\delta(r)$ type singularity. This is so because of the choice of regular gauge when 
describing the point source placed at $r=0$.

Finally one has to consider the limits on the gravitational mass defect. They can be derived from regularity restriction 
for the metric components. In case of $\varrho=0$ the multiplier of the delta function $e^{\varphi(r)}$ in 
(\ref{fnalsystem}) is infinite while in case of $\varrho=1$ the cut $\rho(0)$ becomes infinite. Both of these cases have 
to be excluded from the domain in which $\varrho$ varies because of these two singularities and then the restriction for 
$\varrho$ becomes stronger: $\varrho\in(0,1)$. It is the case of strong gravitational fields that take place when $\varrho$ 
approaches $1$ and the asymptotic behaviour $\varrho\to 1\Rightarrow\rho_0=\infty$ has a simple physical interpretation. It is 
connected to the luminosity of the point source in such case, which tends to zero due to $L\le L_0\sim\rho_0^{-2}$ and thus 
stronger gravitational field corresponds to darker object. This is exactly the same physical situation as in purely 
gravitational point source \cite{fiziev} where $\rho_0=2M/(1-\varrho^2)$ and the difference between the 
two values of $\rho_0$ is
\ben\nonumber
\lim_{\varrho\to 1}\left(\rho_0^{\text{Einstein}}-\rho_0^{\text{Einstein--Maxwell}}\right)=\frac{Q^2}{2M},
\een
where Einstein stays for pure gravitational field and Einsten--Maxwel for the current solution.
\section{Solutions for the Electromagnetic Field}
The gravitational field of the point source is given in (\ref{etal}), (\ref{etae}) and (\ref{etag}) with the use of 
(\ref{mcompts}). Now one has to consider the electromagnetic field which is a solution of the third equation in 
(\ref{fsystem}) with $\varphi_3'(r)=\varphi'(r)e^{-\varphi_1(r)}$. It is easy to check that this equation is equivalent to 
\ben\nonumber \varphi_3'(r)=\varphi_3'(0)e^{\varphi_1(r)-\varphi_1(0)}- \frac{Q}{\bar\rho^2}e^{\varphi_1(r)}\Delta(r). 
\een 
Going back to the electric field $\varphi(r)$ this is transformed to 
\ben\label{help3} \varphi'(r)=\left(Q\Delta(r)-\bar\rho^2\frac{\varphi'(0)}{\varrho^2}\right) \left(\frac{1}{\rho}\right)' 
\een 
and straightforward integration gives the solution for $\varphi(r)$ depending on the initial values $\varphi(0)$ and 
$\varphi'(0)$ 
\ben\label{eminit} \varphi(r)=&\varphi(0)+Q\left(\frac{1}{\rho(r)}-\frac{1}{\rho(0)}\right)\Delta(r)- 
\bar\rho^2\frac{\varphi'(0)}{\varrho^2}\left(\frac{1}{\rho(r)}-\frac{1}{\rho(0)} \right). 
\een

To find the initial values one has to compare (\ref{eminit}) to the well known static electromagnetic field outside the 
source, i.e. for $r>0$: 
\ben \nonumber \left(\varphi(0)+\frac{\bar\rho^2\varphi'(0)}{\rho(0)\varrho^2}\right)- \frac{\bar\rho^2\varphi'(0)}
{\rho(r)\varrho^2}=\frac{Q}{\rho(r)}. \een This is equivalent to two equations for the constants $\varphi(0)$ and 
$\varphi'(0)$ and after solving them and performing the rational transformation (\ref{ratch}) the final expression for 
the electromagnetic field becomes 
\ben\label{emsol} \varphi(\zeta)=\frac{Q\theta(\zeta)}{\rho(\zeta)}-\frac{Q}{\rho(0)}\Delta(\zeta). 
\een

The most significant point about this solution is its finite value for $\zeta=0$ 
\ben \varphi(0)=\text{sign}{(Q)}\frac{1-\sqrt{1-\eta^2(1-\varrho^2)}}{\eta}. 
\een 
This shows how an appropriate regularization of the gravitational field of a point particle suppresses the divergencies 
of the electric field as well. This purely GR phenomena is a regularization at classical level which is impossible to 
derive in terms of Newtonian theory of gravitation. The importance of this result comes from the independence of this 
regularization scheme on Maxwell theory. This gives a chance for regular solutions of the point particle problem in 
generalized theories coupling these two like 5D Kaluza Klein, extra dimensions, etc.

Another interesting feature of this cut for the electromagnetic field comes from the restriction on the luminosity 
variable when the charge is greater than the mass of the source, i.e. when real particles are considered. In this 
case it was shown in the previous section that there is a limiting value for the mass defect parameter 
$\varrho_{\text{lim}}$ and thus $\rho(0)>\rho_{\text{cl}}$ which corresponds to $\varphi(0)<Q/\rho_{\text{cl}}$. 
Thus the electromagnetic field of a test particle with the same charge as the point source is restricted to values 
smaller than the Keplerian mass of the source $M$.
\section{The Final Solution}
In order to arrive at the final form of the point particle solution one has to perform a coordinate transformation 
of the dimensionless quantity $\zeta$ defined in (\ref{ratch}) to a new radial coordinate. The easiest way to do 
this is by introducing a constant $\tilde r$ such that \ben \nonumber\zeta=\frac{r}{\tilde{r}}. \een This is a rational 
transformation which is well defined in the domain $\zeta\in(0,\infty)$ and will not lead to any poles in the solutions 
even when the point $\zeta=0$ is added.

The scaling parameter $\tilde r$ should be such that $g_{\text{tt}}$ has the standard asymptotic behaviour when 
$r\to\infty$: $g_{\text{tt}}\sim 1-\frac{2M}{r}$. This condition gives different expressions for $\tilde r$ in 
each case $\eta<1$, $\eta=1$, $\eta>1$: 
\ben\nonumber \tilde{r}=\frac{\sqrt{1-\eta^2}}{\Delta\Phi^-}M,\,\eta<1;\,\,\, \tilde{r}=\frac{\varrho}{1-\varrho}M,\,\eta=1;
\,\,\, \tilde{r}=\frac{\sqrt{\eta^2-1}}{\Delta\Phi^+}M,\,\eta>1. 
\een 
Using this result and $dr=\frac{r_\infty d\zeta}{(\zeta+1)^2}$ coming from (\ref{ratch}) the point particle solution of 
Einstein-Maxwell equations can be written in the following way 
\ben
ds^2&=&e^{2\varphi_\alpha(r)}\left(dt^2-\frac{dr^2}{N_\alpha^4(r)}\right)-\rho(r)^2\left(d\theta^2+\sin{\theta}^2d\phi^2\right),\\
\nonumber \varphi(r)&=&\frac{Q}{\rho(r)}, 
\een 
where the notation $\varphi_\alpha(r;M,M_0,Q)=\varphi_1(r)$ was used. Each case for $\eta$ leads to different expressions 
for each of the functions $\varphi(r)$ and $\rho(r)$ and there are three different solutions determined by the following functions:

1) In case $M>Q$ 
\ben \varphi_\alpha(r)&=&\ln\left(\frac{\frac{\sqrt{M^2-Q^2}}{r+\tilde r_1}}{\sinh{\left(\frac{\sqrt{M^2-Q^2}}{r+\tilde r_1}\right)}}
\frac{r+\tilde r_1}{\rho(r)}\right),\\
\nonumber \rho(r)&=&M+\sqrt{M^2-Q^2}\coth\left(\frac{\sqrt{M^2-Q^2}}{r+\tilde r_1}\right)+ 
M_0\frac{\sinh{\left(\frac{\sqrt{M^2-Q^2}}{r+\tilde r_1}\frac{r}{\tilde r_1}\right)}} 
{\sinh{\left(\frac{\sqrt{M^2-Q^2}}{r+\tilde r_1}\right)}}\Delta(r), 
\een 
where the scaling parameter $\tilde r_1$ depends on $M,M_0$ and $Q$: 
\ben \nonumber \tilde r_1:=\frac{\sqrt{M^2-Q^2}}{\text{arcsinh}\left(\frac{\sqrt{M^2-Q^2}}{Q}\frac{M_0}{M}\right)- \text{arcsinh}
\left(\frac{\sqrt{M^2-Q^2}}{Q}\right)}. 
\een

2) In case $Q=M$ 
\ben \nonumber
\varphi_\alpha(r)&=&\ln\left(\frac{r+\tilde r_2}{\rho(r)}\right),\\
\rho(r)&=&M+\tilde r_2+r+M_0\frac{r}{\tilde r_2}\Delta(r),\\
\nonumber \tilde r_2&=&\frac{M^2}{M_0-M}. 
\een

3) In case $M<Q$ 
\ben \varphi_\alpha(r)&=&\ln\left(\frac{\frac{\sqrt{Q^2-M^2}}{r+\tilde r_3}}{\sin{\left(\frac{\sqrt{Q^2-M^2}}{r+\tilde r_3}\right)}}
\frac{r+\tilde r_3}{\rho(r)}\right),\\
\nonumber \rho(r)&=&M+\sqrt{Q^2-M^2}\cot\left(\frac{\sqrt{Q^2-M^2}}{r+\tilde r_3}\right)+ 
M_0\frac{\sin{\left(\frac{\sqrt{Q^2-M^2}}{r+\tilde r_3}\frac{r}{\tilde r_3}\right)}} 
{\sin{\left(\frac{\sqrt{Q^2-M^2}}{r+\tilde r_3}\right)}}\Delta(r), 
\een 
where a different scaling parameter $\tilde r_3$ is used: 
\ben \nonumber \tilde r_3:=\frac{\sqrt{Q^2-M^2}}{\arcsin\left(\frac{\sqrt{Q^2-M^2}}{Q}\frac{M_0}{M}\right)- 
\arcsin\left(\frac{\sqrt{Q^2-M^2}}{Q}\right)}. 
\een 
In this case there is also a restriction for the charge of the point particle: 
\ben \nonumber Q\le\frac{M}{\sqrt{1-\left(\frac{M}{M_0}\right)^2}} 
\een 
providing that the minimal value of $\rho(r)$ is the particle's classical radius.

The function $N_\alpha(r)$ in each case is given by 
\ben N_\alpha(r)=\frac{M}{\tilde r}\frac{(r+\tilde r)}{\rho(r)} 
\een 
in terms of $\rho(r)$ and $\tilde r=\tilde r_{1,2,3}$.

All these functions are defined in the interval $r\in(0,\infty)$ and are singularity free everywhere in 
this domain. A significant property of the considered solution is its regularity even when $r\to 0$ which 
is very uncommon due to the well known divergences emerging in the classical theory.

The functions $\varphi_\alpha(r)$ and $\varphi(r)$ with the use of $\rho(r)$ in each case correspond to corrected 
Newton and Coulomb potentials of the point source respectively. In order to investigate the obtained corrections 
it is useful to rewrite the scaling parameter for each case in terms of $\rho_0=\rho(0)=Mz(0)$ (see (\ref{zzero}) 
for the definition of $z(0)$):
\ben\nonumber
\tilde r_1=\frac{\sqrt{M^2-Q^2}}{\text{arctanh}\frac{\sqrt{M^2-Q^2}}{\rho_0-M}},\,\,
\tilde r_2=\rho_0-M,\,\,
\tilde r_3=\frac{\sqrt{Q^2-M^2}}{\arctan\frac{\sqrt{Q^2-M^2}}{\rho_0-M}}.
\een
Now using these one can consider the following asymptotic cases:
\begin{itemize}
\item{Vanishing electric charge $Q\to 0$ and thus $Q\ll M$. This case leads to the pure gravitational point source solution 
derived in \cite{fiziev}:}
\ben\nonumber
\rho_0\to\frac{2M}{1-\varrho^2}\Rightarrow\tilde r_1\to\frac{M}{\ln{M_0/M}},
\een
and the corresponding corrected Newton and Coulomb potentials are
\ben
\varphi_\alpha(r)\approx -\frac{M}{r+\frac{M}{\ln{M_0/M}}},\,\, \varphi(r)\approx\left(1-e^{-\frac{2M}{r+\frac{M}{\ln{M_0/M}}}}
\right)\frac{Q}{M}
\een
\item{Mass of the point source $M\ll Q$ where the condition $\eta^{-2}+\varrho^2\ge 1$ is satisfied. This case leads to the 
following asymptotic behaviour of the scaling parameter $\tilde r_3$:}
\ben\nonumber
\rho_0\to\frac{Q^2}{M}\Rightarrow\tilde r_3\to\frac{Q^2}{M},
\een
where the limit is considered over the the curve $\eta^{-2}+\varrho^2=1$. The corresponding corrected Newton and Coulomb 
potentials are
\ben
\varphi_\alpha(r)\approx -\ln{\left(\cos\frac{Q}{r+\frac{Q^2}{M}}\right)},\,\, \varphi(r)\approx\tan{\frac{Q}{r+\frac{Q^2}{M}}}
\een
\end{itemize}

\section{Hamiltonian and Energy of the Fields in BRG}
After solving the field equations using the lagrangian formalism one can step back to the beginning and derive a 
hamiltonian density corresponding to (\ref{totlagr}). In the scope of the present article this will be used to find 
the total energy of the charged point source. This procedure was used in \cite{fiziev} in case of neutral point 
source and will be repeated here. If BRG is taken into account the total lagrangian (\ref{totlagr}) depends on three 
``dynamical'' fields $\varphi_1$, $\varphi_2$, $\varphi_3$ and one can obtain their conjugate momenta from 
$\pi_i:=\frac{\partial\mathcal{L}_{\text{tot}}}{\partial\varphi_i'}$: 
\ben \pi_1=-\bar\rho^2\varphi_1',\, \, \pi_2=\bar\rho^2\varphi_2',\, \, \pi_3=\bar\rho^2e^{-2\varphi_1}\varphi'. 
\een 
Using these relations the total hamiltonian is simply 
\ben\nonumber \mathcal{H}_{\text{tot}}=\sum_{i}\pi_i\varphi_i-\mathcal{L}_{\text{tot}}(\pi,\varphi)= 
\varepsilon_{\text{tot}}+\left(M_0e^{\varphi_1}+Q\varphi\right)\delta(r), 
\een 
where (\ref{epsilons}) was used. The total energy is an integral of this hamiltonian density over the whole domain of 
the radial coordinate $r$: 
\ben\nonumber E_{\text{tot}}:=\int_0^{\infty}\mathcal{H}_{\text{tot}}(r)dr=M+M\left[1- \sqrt{1-(1-\varrho^2)\eta^2}\right]. 
\een 
The second part of this expression comes from the fact that $\varepsilon_{\text{tot}}=\varepsilon_1+\epsilon_2+\varepsilon_3$ 
satisfies the third relation in (\ref{firstints}).

Next it is useful to separate the energies corresponding to electromagnetic and gravitational field. The easiest way to do 
this is using the energy density of the electromagnetic field derived from its energy-momentum tensor: 
\ben\label{emdensity} \epsilon_{\text{EM}}={T_{\text{EM}}}_{0}^{0}=\frac{Q^2}{8\pi\rho^4}. 
\een 
The electromagnetic part of the total energy is then determined as 
\ben E_{\text{EM}}=\int_{\rho_0}^{\infty}4\pi\rho^2\epsilon_{\text{EM}}d\rho=\frac{Q^2}{\rho_0}= 
M\left(1-\sqrt{1-(1-\varrho^2)\eta^2}\right). 
\een

The energy of the gravitational field can be obtained from the relation \cite{fiziev} 
$E_{\text{tot}}=M_0+E_{\text{GR}}+E_{\text{EM}}$ so that 
\ben \nonumber E_{\text{GR}}&:=M-M_0=-(1-\varrho)M_0<0. 
\een 
This result was recently obtained for the pure gravitational field of a point source without electric charge in 
\cite{fiziev} and it turns out that the charge does not change it at all..

\section{Integral Theorems and Invariant Quantities}
After fixing the gauge condition (\ref{brg}) the two field equations containing delta functions are 
\ben\label{help2} \nonumber
&&\bar\Delta_r\varphi_1=\varphi_3'^2+\frac{M_0}{\bar\rho^2}e^{\varphi_1}\delta(r),\\
&&\bar\Delta_r\varphi_3=\varphi_1'\varphi_3'-\frac{Q}{\bar\rho^2} e^{\varphi_1}\delta(r). 
\een 
These can be used to derive integral invariants for the fields just like the Gauss law in case of electromagnetic 
field only. Actually one of these integral invariants will be just the Gauss law with a correction due to General 
Relativity. In order to derive these invariants one has to transform (\ref{help2}) to equations for $\varphi_1$ and 
$\varphi$ corresponding to the gravitational and electromagnetic field respectively. After having this done the equations read 
\ben \nonumber &&\Delta_r\varphi_1=\frac{\varphi'^2}{-g_{\text{rr}}g_\text{tt}}+
\frac{M_0}{\bar\rho^2}\frac{\sqrt{g_{\text{tt}}}}{-g_{\text{rr}}}\delta(r),\\
\nonumber &&\Delta_r\varphi=\frac{2\left(\sqrt{g_\text{tt}}\right)'\varphi'}{-g_{\text{rr}} \sqrt{g_\text{tt}}}- 
\frac{Q}{\bar\rho^2}\frac{g_{\text{tt}}}{-g_{\text{rr}}}\delta(r). 
\een

A combination of (\ref{help3}) and (\ref{newheav}) as well as (\ref{mcompts}) in BRG give 
$\varphi'=-Q\theta(r)\left(\frac{1}{\rho}\right)'$ and $\frac{\rho^2\sqrt{g_{\text{tt}}}}{\bar\rho^2\sqrt{-g_{\text{rr}}}}=1$ 
and using these the field equations are transformed to the following 
\ben \nonumber &&\Delta_r\varphi_1=\frac{1}{\sqrt{|g^{(3)}|}}\left[\frac{Q^2\theta(r)}{\bar\rho^2}
\sqrt{g_{\text{tt}}}+M_0\delta(r)\right],\\
\nonumber &&\Delta_r\varphi=\frac{1}{\sqrt{|g^{(3)}|}}\left[2\left(\sqrt{g_{\text{tt}}}\right)' \theta(r)-
\sqrt{g_{\text{tt}}}\delta(r)\right]Q. \een Due to $r\in[0,\infty)$ this reads \ben \nonumber &&
4\pi\sqrt{|g^{(3)}|}\Delta_r\varphi_1=4\pi\left[8\pi\left(\frac{Q^2}{8\pi\rho^4}\right) \sqrt{|g^{(3)}|}+
M_0\delta(r)\right],\\
\nonumber &&4\pi\sqrt{|g^{(3)}|}\Delta_r\varphi=4\pi Q\left[2\left(\sqrt{g_{\text{tt}}}\right)'-
\sqrt{g_{\text{tt}}}\delta(r)\right]. 
\een

Integration of these expressions over $r$ is equivalent to integration of the left hand side over the invariant three 
dimensional volume and the final result is: 
\ben \nonumber &&\int_{\mathcal{M}^3}d^3\text{\bf r}\, \sqrt{|g^{(3)}|}\left[\Delta(\ln{
\sqrt{g_{\text{tt}}}})-8\pi\epsilon_{EM}\right]=4\pi M_0,\\
\nonumber &&\int_{\mathcal{M}^3}d^3\text{\bf r }\, \sqrt{|g^{(3)}|}\Delta\varphi=4\pi Q\left(2-3\varrho\right), 
\een 
where the notation (\ref{emdensity}) was used. The second integral here correspond to the well known Gauss law in 
electrodynamics. In this sense the regularization of the gravitational field leads to a change in the Gauss law for 
the electromagnetic field besides the above mentioned regularization.

Another interesting result can be obtained from integration of the two differential equations multiplied by a factor 
$\sqrt{g_{\text{tt}}}$ which gives two other invariants: 
\ben \nonumber &&\int_{\mathcal{M}^3}d^3\text{\bf r}\, \sqrt{|g^{(4)}|}\left[\Delta(\ln{
\sqrt{g_{\text{tt}}}})-8\pi\epsilon_{EM}\right]=4\pi M,\\
\nonumber &&\int_{\mathcal{M}^3}d^3\text{\bf r }\, \sqrt{|g^{(4)}|}\Delta\varphi=4\pi Q(1-2\varrho). 
\een

It is obvious now that one ends up with four equations for the parameters $M,\, Q\, M_0$ and $\varrho$ which can be 
solved and the latter can be expressed in terms of the four invariant integrals.
\section{Conclusion}
It was shown that when regular gauge conditions take place in the charged point source problem the solutions for 
both gravitational and electromagnetic fields are finite and have their regular behaviour at $r=0$. This is a fact 
coming from the nontrivial topology of spacetime induced by the regularity of the gauge conditions. It is interesting to note that 
this topology is strongly connected to the definition of a physical point source described in the very beginning. It 
describes the assumption for finite structure of the point in a quite natural way. In other words when the condition 
$L\gg a$ discussed in the beginning is no more satisfied the geometry has its own way of protecting the source's 
structure which is quite different from the mathematical one. This way leads to finite luminosity as well as finite 
field potentials, $\varphi(r)$ and the metric components $\varphi_{1,2}(r)$, at the point where the source is placed. 
The regular behaviour of the fields is described by corrections to Newton's and Coulomb's laws at a classical level 
which take place only at distances defined by a new scaling parameter $\tilde r$. In this way inclusion of the 
gravitational field leads to a natural regularization at a classical level. This is a fact expected by many of the 
people creating GR in its beginning but the nontrivial geometry leading to such result is the price one has to pay. 
This geometry is essentially different from the geometry used in \cite{ADM} to describe charged point particles due to 
the difference between the Keplerian and bare mass of the particle.

Another consequence of these geometrical properties is the correction in Gauss theorem when considering the 
electromagnetic field. This correction comes from a new parameter, recently used in description of GR problems, 
which is the gravitational mass ratio $\varrho$. This is a parameter defining the geometrical behaviour and the new 
distance scale of the solution and has a clear physical meaning -- it describes the difference between the bare and 
Keplerian mass of an object. It is a very important quantity in this problem and there are several recent attempts 
to measure its value from supernovas which will provide some evidential reasons for its use. 

Still after the derivation of the pure gravitational solution in \cite{fiziev} and the one for a charged point source 
in this article many open questions rise. The most important ones are the behaviour of test particles and classical fields 
and the corresponding scattering problem, geodesic completeness and mostly the derivation of these solutions as a limit from 
finite objects. All these are subject to future development in this direction.

\end{document}